\documentclass[prb, showpacs]{revtex4}
\usepackage{amssymb}
\usepackage{bm}
\usepackage{graphicx}%

\begin{document}
\title
{Dynamic equation for quantum Hall bilayers with spontaneous
interlayer coherence: The low-density limit}

\author{A.\,I.\,Bezuglyj}

\email{bezugly@ic.kharkov.ua}

\affiliation{%
NSC Kharkov Institute of Physics and Technology, 61108, Kharkov, Ukraine}

\author{S.\,I.\,Shevchenko}

\email{shevchenko@ilt.kharkov.ua}
\affiliation{%
B.\,I.\, Verkin Institute for Low Temperature Physics and Engineering, 61103, Kharkov, Ukraine}

\begin{abstract}
  The bilayer systems exhibit the Bose-Einstein condensation of
excitons that emerge due to Coulomb pairing of electrons belonging to one layer with the holes belonging to the
other layer. Here we present the microscopic derivation of the dynamic equation for the condensate wave function
at a low density of electron-hole ($e-h$) pairs in  a strong magnetic field perpendicular to the layers and an
electric field directed along the layers. From this equation we obtain the dispersion law for collective
excitations of the condensate and calculate the electric charge of the vortex in the exciton condensate. The
critical interlayer spacing, the excess of which leads to a collapse of the superfluid state, is estimated. In
bilayer systems with curved conducting layers, the effective mass of the $e-h$ pair becomes the function of the
$e-h$ pair coordinates, the regions arise, where the energy of the $e-h$ pair is lowered (exciton traps), and
lastly $e-h$ pairs can gain the polarization in the basal plane. This polarization leads to the appearance of
quantized vortices even at zero temperature.

\end{abstract}

\pacs{73.21.Fg, 71.35.Ji}

\maketitle

\section{INTRODUCTION}
In recent years considerable attention has been focused on the so-called bilayers, i.e. the systems that
represent two thin and closely-spaced conducting layers separated by the energy barrier for carriers \cite{1a}.
This has been mainly caused by a successful realization of such systems in GaAs/AlGaAs heterostructures. At
small interlayer spacings the Coulomb interaction of carriers belonging to adjacent layers proves to be rather
essential. The interaction as well as a two-dimensional character of carrier motion determine the unique
properties of bilayers. In particular, in the bilayer system there may arise a collective state, which is
characterized by the interlayer phase coherence and, as a consequence, the system becomes superfluid.

One of the examples of the bilayer system is the $n-p$ system
consisting of the $n$-layer with electron conductivity and the
$p$-layer with hole conductivity. In the development of the ideas
by Keldysh and collaborators \cite{7} it has been established that
in this system in the weak-coupling limit the attraction of
electrons from the $n$-layer conduction band and holes from the
$p$-layer valence band  gives rise to the coherent BCS-type state
of electron-hole ($e-h$) pairs \cite{1,2}. By virtue of the
spatial separation of pair components, the superfluid motion of
$e-h$ pairs reveals itself in an emergency of equal and opposite
directed dissipationless electric currents, those may be detected
in the experiment.

The pairing of electrons and holes as well as the condensation of $e-h$ pairs is favored in a strong magnetic
field $H$, normal to the layers, provided that the magnetic length $l_H$ $(l_H=\sqrt{\hbar c/eH})$ is
considerably shorter than the Bohr radii of both the electron and the hole, and the carriers in the both layers
occupy the states in the lowest Landau level \cite{3}. In this case (in the symmetric gauge), the wave functions
of both the electron and the hole are "localized" in the region of about $l_H$  in size (rather than the size of
the effective Bohr radius), that substantially increases the binding energy of the $e-h$ pair at small
interlayer spacings ($d < l_H$). Note that at low pair densities, $n_p\ll l_H^{-2}$, the electron-hole pairing
occurs in the real space (strong-coupling limit of $e-h$ pairs) rather than in the momentum space. The
properties of an individual $e-h$ pair in the two-layer system (indirect magnetic exciton) have been considered
in Ref.6.

Another example of the phase-coherent system is the $n-n$ system
including two electron layers in the magnetic field perpendicular
to the layers with the total filling factor $\nu_T= \nu_1 + \nu_2
= 1$. Despite the apparent difference between the $n-n$ and $n-p$
systems, they are, in essence, fully similar. Indeed, at the
equality $\nu_T= 1$, the number of electrons in the first layer is
equal to the number of free states in the lowest Landau level in
the second layer (for certainty, we assume $\nu_1 < \nu_2 $). Then
the transition to the "hole" representation in the second layer
\cite{6} gives the one-to-one correspondence between the $n-n$
system and the above-discussed $n-p$ system in the magnetic field.
So, in the systems of both types the interlayer phase coherence is
the consequence of Bose condensation of $e-h$ pairs, and due to
this reason the superfluidity in bilayer systems is often called
the "exciton superfluidity".

As mentioned above, the exciton superfluidity phenomenon was
predicted for the $n-p$ systems in the weak-coupling limit and at
a high density of $e-h$ pairs rather long ago \cite{1,2}. At the
present time, one has impressive evidence that the superfluidity
in question was revealed experimentally in quantum Hall $n-n$
systems at $\nu_1 = \nu_2 =1/2 $ \cite{8,9}. This experimental
situation corresponds to the intermediate pair density, when the
distance between the pairs is of order of the pair size. We stress
that the $\nu_1 = \nu_2 =1/2 $ condition is not necessary: for the
exciton superfluidity the equality between electron densities and
hole densities, i.e., $\nu_T= 1$, is necessary.

The experimental revealing of exciton superfluidity in bilayer
systems is a rather strong incentive to its further studies. The
study of superfluidity  of strongly bound $e-h$ pairs in the
low-density limit is of particular interest from the viewpoint of
theory, because one can expect that the $e-h$ pairs with spatially
separated components behave as a weakly nonideal Bose gas
\cite{9a}, which is described by a relatively simple differential
equation of the Gross-Pitaevskii type \cite{9b}. In the present
paper  in the mean field approximation we derive microscopically
this equation for the   condensate wave function. We consider the
low-density limit of $e-h$ pairs in quantum Hall $n-n$ systems
with $\nu_T =1$. It is assumed that the magnetic length is
considerably shorter than the Bohr radii of electrons and holes.
Besides, it is considered that the interlayer tunneling is
negligibly small. The equation is applied to obtain the dispersion
law of collective excitations and electric charge of the vortex in
the condensate of electrically neutral $e-h$ pairs.

\section{THE DYNAMIC EQUATION}

We now proceed to the derivation of the dynamic equation. The Hamiltonian of the bilayer electron-hole system in
the perpendicular magnetic field is

\begin{equation}\label{1}
\hat H=\int d{\bf r}\sum_k \psi_k^+({\bf r}) \hat h_k({\bf r})
\psi_k({\bf r}) + {1\over2} \int d{\bf r} d{\bf r'}\sum_{k,l}
V_{kl} ({\bf r}-{\bf r'}) \psi_k^+({\bf r}) \psi_l^+({\bf r'})
\psi_l({\bf r'}) \psi_k({\bf r})  .
 \end{equation}
Here and further on it is assumed that the indices $k$, $l$ $= 1
(2)$ indicate the electron (hole) layer. The operators
$\psi_k^+({\bf r})$ and $\psi_k({\bf r})$ are the operators of
electron (hole) creation and annihilation at the point with
two-dimensional radius-vector ${\bf r}$. The spin indices are
omitted, since electrons are considered spin-polarized. The
kinetic energy operator has the form

\begin{equation}\label{2}
\hat h_k({\bf r})={1\over{2m_k}}\Bigl [i\hbar \frac{\partial}{\partial{\bf r}}
+(-1)^k {e\over c}{\bf A}_k ({\bf r})\Bigr ]^2,
 \end{equation}
and the Coulomb interaction energy of carriers is given by

\begin{equation}\label{3}
V_{kl} ({\bf r}-{\bf r'}) = (-1)^{k+l}{e^2\over{\varepsilon
\sqrt{({\bf r}-{\bf r'})^2 + (1-\delta_{k,l})d^2}}}.
 \end{equation}
In formulas (\ref{2}), (\ref{3}) $m_k$ is the electron mass $(k=1)$ or the hole mass
$(k=2)$, $\varepsilon $ is the dielectric constant, $d$ is the interlayer spacing,
$\delta_{k,l}$ is the Kronecker delta-symbol.

To derive the dynamic equation for the condensate wave function we
shall follow Keldysh's paper \cite{10}, according to which the
coherent state of $e-h$ pairs, in the mean field approximation,
may be represented by the vector $|\phi\rangle = \hat D_\phi
|0\rangle$, where the unitary operator $\hat D_\phi$ is written as

\begin{equation}\label{4}
\hat D_\phi = \exp\Bigl\{\int d{\bf r}_1 d{\bf r}_2 [\psi_1^+({\bf r}_1) \Phi({\bf r}_1,{\bf r}_2,t)e^{-i\mu
t}\psi_2^+({\bf r}_2)- \psi_2({\bf r}_2)\Phi^*({\bf r}_1,{\bf r}_2,t)e^{i\mu t}\psi_1({\bf r}_1)] \Bigr\}.
\end{equation}
The vector $|0\rangle$ is the vacuum state of the system: $\psi_k |0\rangle=0$. The
unknown function $\Phi({\bf r}_1,{\bf r}_2,t) $ and the chemical potential $\mu$ of
$e-h$ pairs, which enter into $\hat D_\phi$, should be found from the Schrodinger
equation for the vector $|\phi\rangle$ . This equation is conveniently written as

\begin{equation}\label{5}
(i\hbar \hat D_\phi^+\frac{\partial}{\partial t}\hat D_\phi -
\hat D_\phi^+H\hat D_\phi)|0\rangle=0.
\end{equation}
In the expression $\hat D_\phi^+H\hat D_\phi$ the operators $\hat D_\phi^+$ and $\hat
D_\phi$ realize the linear transformation of the creation/annihilation operators of
electrons and holes. For example, the operator $\psi_1({\bf r})$ is transformed as

$$\hat D_\phi^+\psi_1({\bf r})\hat D_\phi= \int d{\bf r}' [C({\bf r},{\bf
r}')\psi_1({\bf r}') + S({\bf r},{\bf r}') \psi_2^+({\bf r}')e^{-i\mu t}],$$ where
with an accuracy up to terms cubed in $\Phi$ we have

$$C({\bf r},{\bf r}')=\delta ({\bf r}-{\bf r}')-
(1/2)\int d{\bf r}''\Phi({\bf r},{\bf r}'')\Phi^*({\bf r}',{\bf r}'');$$

$$S({\bf r},{\bf r}')=\Phi({\bf r},{\bf r}')- (1/6)\int\int d{\bf r}''d{\bf
r}'''\Phi({\bf r},{\bf r}'') \Phi^*({\bf r}''',{\bf r}'')\Phi({\bf r}''',{\bf r}').$$ As a result of
transformations, Eq.(\ref{5}) in the Hartree-Fock approximation takes the following form:

\begin{equation}\label{6}
\int d{\bf r}_1 d{\bf r}_2 \Bigl [\sum_k\psi_k^+({\bf r}_1) \tilde h_k({\bf r}_1,{\bf r}_2,t)\psi_k({\bf r}_2) +
\psi_1^+({\bf r}_1)Q({\bf r}_1,{\bf r}_2,t)e^{-i\mu t}\psi_2^+({\bf r}_2) + \psi_2({\bf r}_2)Q^*({\bf r}_1,{\bf
r}_2,t)e^{i\mu t}\psi_1({\bf r}_1) \Bigr ]|0\rangle=0.
\end{equation}
We do not give here the explicit expressions for $\tilde h_k({\bf r}_1,{\bf r}_2,t)$  and $Q({\bf r}_1,{\bf
r}_2,t)$ because they are too cumbersome (for more details, see Ref. 12). Since $\psi_k |0\rangle=0$, equality
(\ref{6}) is equivalent to the equality $Q({\bf r}_1,{\bf r}_2,t)=0$, that represents a nonlinear
integro-differential equation for the function $\Phi({\bf r}_1,{\bf r}_2,t) $. In the general case, $Q$
comprises the convolutions over the coordinates of an arbitrary large odd number of functions $\Phi$. However,
in what follows we shall retain the terms not higher than the third order in $\Phi$, because we are interested
in the low-density limit of $e-h$ pairs. We first consider the non-interacting pairs. For this purpose, in the
equation $Q({\bf r}_1,{\bf r}_2,t)=0$ we keep only the linear in $\Phi$ terms. We obtain

\begin{equation}\label{7}
i\hbar\frac{\partial\Phi}{\partial t} = [\hat h_1({\bf r}_1) +
\hat h_2({\bf r}_2)+V_{12} ({\bf r}_1-{\bf r}_2)]\Phi - \mu\Phi.
\end{equation}
Equation (\ref{7}) can be transformed into the equation for the wave function of the $e-h$ pair condensate
(slowly varying in space and time) by averaging it over "fast" variables that describe the intrinsic degrees of
freedom of $e-h$ pairs. To separate the intrinsic degrees of freedom, we turn from the electron/hole coordinates
${\bf r}_1$ and ${\bf r}_2$ to the difference coordinate $\tilde{\bf r} = {\bf r}_1 - {\bf r}_2$ and to the
coordinate of the center of inertia of the pair ${\bf R} = (m_1{\bf r}_1+m_2{\bf r}_2)/(m_1+m_2)$. In the
symmetric gauge ${\bf A}_k({\bf r}_k)=(1/2){\bf H}\times{\bf r}_k$ , it is convenient to seek the approximate
solution of Eq.(\ref{7}) for the states belonging to the lowest branch of the exciton spectrum in the following
form:

\begin{equation}\label{8}
\Phi( \tilde{\bf r},{\bf R},t)=\exp\Bigl({{ie}\over{2\hbar c}}{\bf R}\cdot ({\bf H}\times\tilde{\bf r})+
{\gamma\over 2}\tilde{\bf r}\cdot \frac{\partial}{\partial{\bf R}}\Bigr) \varphi_0(\tilde{\bf r}-\hat{\bf
\rho})\Psi({\bf R},t).
\end{equation}
Here $\gamma=(m_2-m_1)/(m_2+m_1)$, $\varphi_0({\bf r})= (\sqrt{2\pi}l_H)^{-1}\exp(-r^2/4l_H^2)$ is the wave
function of the state at the lowest Landau level, and $\hat{\bf\rho}= -i{{\hbar c}\over{eH^2}}({\bf
H}\times\frac{\partial}{\partial{\bf R}})$. The structure of expression (\ref{8}) is determined by the solution
of the steady-state Schrodinger equation for the $e-h$ pair in the bilayer system. Indeed, at $\Psi({\bf R})
\propto \exp({i\over\hbar} {\bf p\cdot R})$ expression (\ref{8}) turns into the wave function of the exciton
with the momentum equals to ${\bf p}$ \cite{11,4}. In the case of a slowly varying function $\Psi({\bf R})$
expression (\ref{8}) presents a wave packet composed of low-momenta exciton states. To obtain the sought-for
equation for $\Psi({\bf R},t)$, we act upon Eq.(\ref{7}) by the operator $\varphi_0(\tilde{\bf r}-\hat{\bf\rho})
\exp\Bigl(-{{ie}\over{2\hbar c}}{\bf R}\cdot({\bf H}\times\tilde{\bf r})- {\gamma\over 2}\tilde{\bf
r}\cdot\frac{\partial}{\partial{\bf R}}\Bigr)$ and integrate over the variable $\tilde{\bf r}$ related to the
intrinsic degrees of freedom of the $e-h$ pair. As a result of integration, the first two terms in the
right-hand side of Eq.(\ref{7}) give $(\hbar\omega_c/2)\Psi({\bf R},t)$, where $\omega_c=({1\over m_1}+ {1\over
m_2}){eH\over c}$. Since ${\bf R}$ is the "slow" variable, the integral of the Coulomb term in Eq.(\ref{7}) can
be calculated by expanding $\varphi_0(\tilde{\bf r}-\hat{\bf\rho})$ in powers $\hat{\bf\rho}$ to an accuracy of
the second-order terms. Putting the chemical potential equal to the ground-state energy of noninteracting pairs,
$\mu = (\hbar\omega_c/2) +E_0$, where the binding energy of the $e-h$ pair is given by

\begin{equation}\label{9}
E_0=-{e^2\over\varepsilon l_H} \sqrt{{\pi\over 2}}
\exp\Bigl({d^2\over 2l_H^2}\Bigr)
{\rm erfc}\Bigl({d\over \sqrt2 l_H}\Bigr),
\end{equation}
Eq.(\ref{7}) can be transformed into the Schrodinger equation for the wave function of the condensate of
noninteracting $e-h$ pairs:

\begin{equation}\label{9a}
i\hbar\frac{\partial\Psi({\bf R},t)}{\partial t} = -{\hbar^2\over 2M_H}
\frac{\partial^2}{\partial{\bf R}^2}\Psi({\bf R},t).
\end{equation}
In the case considered (magnetic length $l_H$  is considerably
shorter than the Bohr radii of both the electron and the hole),
the effective mass $M_H$ of the $e-h$ pair arises due to the
Coulomb interaction between the electron and the hole, and is
independent of their masses $m_1$ and $m_2$ \cite{11,4}:

\begin{equation}\label{10}
M_H = {2\varepsilon\hbar^2\over e^2l_H}\sqrt{{2\over\pi}}
\Bigl[\Bigl(1+{d^2\over l_H^2}\Bigr)\exp\Bigl({d^2\over 2l_H^2}\Bigr)
{\rm erfc}\Bigl({d\over \sqrt2 l_H}\Bigr)-\sqrt{{2\over\pi}}
{d\over l_H}\Bigr]^{-1}.
\end{equation}
The mass $M_H$ increases with both the increasing magnetic field
and the increasing spacing between the layers.

To take into account the interaction between $e-h$ pairs, it is
necessary to include the cubic terms in the equation $Q({\bf
r}_1,{\bf r}_2,t)=0$ apart from the linear in $\Phi$ terms. At a
low density of $e-h$ pairs the contribution from these terms can
also be calculated with the use of the perturbation theory. The
correction to the chemical potential of the $e-h$ pair is equal to

\begin{equation}\label{11}
\mu^{(1)}={e^2 \over \varepsilon}\Bigl[4\pi d -
(2\pi)^{3/2}l_H +(2\pi)^{3/2}l_H \exp\Bigl({d^2\over 2l_H^2}\Bigr)
{\rm erfc}\Bigl({d\over \sqrt2 l_H}\Bigr)\Bigr]n_p
\end{equation}
In Eq.(\ref{11}) the first term is related to the energy of the electric field between the electron layer and
the hole layer, while the second and third terms describe the exchange contributions from the intralayer
interaction and interlayer interaction, respectively \cite{12,13}. At $d\ll l_H$ we have
$\mu^{(1)}=\sqrt{2}\pi^{3/2}{e^2 d^2 \over \varepsilon l_H}n_p$. Note that the interaction between $e-h$ pairs
falls off as the distance between the layers decreases, and at $d=0$ the $e-h$ pairs form the ideal gas
\cite{14}. (The last statement is valid under the lowest Landau level approximation).

The nonlinear (cubic in $\Psi$) term in the sought-for equation is calculated by substitution of expression
(\ref{8}) (without derivatives with respect to the "slow" variable ${\bf R}$) into cubic in $\Phi$ terms of the
equation $Q({\bf r}_1,{\bf r}_2,t)=0$. Then the result should be multiplied by $\varphi_0(\tilde{\bf r})
\exp\Bigl(-{{ie}\over{2\hbar c}}{\bf R}\cdot({\bf H}\times\tilde{\bf r})\Bigr)$ and integrated over $\tilde{\bf
r}$. As a result we arrive at the following nonlinear equation for the wave function of the $e-h$ pair
condensate:

\begin{equation}\label{12}
i\hbar\frac{\partial\Psi({\bf R},t)}{\partial t} = -{\hbar^2\over 2M_H}
\frac{\partial^2\Psi({\bf R},t)}{\partial{\bf R}^2}
-{\mu^{(1)}\over n_p}[n_p-|\Psi({\bf R},t)|^2]\Psi({\bf R},t),
\end{equation}
This is nothing but the Gross-Pitaevskii equation in the absence of external fields. From Eq.(\ref{12}) one can
find the spectrum of elementary excitations of the $e-h$ pair condensate. It has the well-known form obtained by
Bogolyubov for a weakly nonideal Bose gas:

\begin{equation}\label{12a}
\omega_B(k) = \sqrt{\Bigl({{\hbar^2k^2}\over {2M_H}}+
2\mu^{(1)}\Bigr){k^2\over {2M_H}}},
\end{equation}
At low $k$ and $d\ne 0$ the dispersion law is acoustic one,
$\omega = c_0 k$, where the speed of sound is
$c_0=\sqrt{\mu^{(1)}/M_H}$. The decrease of the interlayer spacing
$d$ leads to the decrease in $c_0$. In the $d= 0$ case,
$\mu^{(1)}=0$ and spectrum (\ref{12a}) become quadratic in $k$.
The last, as is known, means the absence of superfluidity.

From the coefficients entering into Eq.(\ref{12}) we can form the parameter $\xi = \hbar/\sqrt{2M_H \mu^{(1)}}$,
which has the dimensionality of length. This parameter by its sense coincides with the coherence length in the
superconductivity theory and sets the characteristic scale of wave function module variation. It is not
difficult to see that $\xi \sim l_p(l_H/d)$, where $l_p$ is the average distance between the pairs ($l_p \sim
n_p^{-1/2}$). Hence it follows that at $d\ll l_H$ the ratio $(\xi/l_p) \gg 1$. In this case, a great number of
pairs is situated on the scale $\xi$, that justifies the macroscopic description of the system by the wave
function $\Psi({\bf R})$. The coherence length $\xi$ becomes of the order of the mean distance between the pairs
$l_p$, when $d \sim l_H$, and at $d\gg l_H$ the $\xi$ value (formally) becomes much less than $l_p$. This means
that in the last case the wave function $\Psi({\bf R})$ loses its macroscopic nature. In other words, there must
exist the critical value of the interlayer distance $d_c\sim l_H$, with the excess of which the condition $\xi
\gg l_p$  is violated, and it may be expected that at $d>d_c$ the Bose condensate collapses.

Equation (\ref{12}) can be readily generalized for the case of the electric field ${\bf E}$ parallel to the
layers, provided that ${\bf E}$ weakly varies on the scale $l_H$. To do this, the potential energies of the
electron and the hole (namely: $\sum_k \int d{\bf r}(-1)^k e V({\bf r})\psi_k^+({\bf r}) \psi_k({\bf r})$, where
$V({\bf r})$ is the scalar potential of the electric field) must be included in the Hamiltonian (\ref{1}). In
this case, the addition $eV({\bf r}_2)-eV({\bf r}_1)$ appears in the square brackets of Eq.(\ref{7}). Then it is
necessary to pass on to the coordinates ${\bf R}$ and $\tilde{\bf r}$, introduced before, and to exclude the
fast variable $\tilde{\bf r}$ by means of the above-described procedure. As a result, we come to the following
equation

\begin{equation}\label{13}
i\hbar\frac{\partial\Psi({\bf R},t)}{\partial t} = \Bigl\{{1\over 2M_H} \Bigl(i\hbar
\frac{\partial}{\partial{\bf R}}-{\alpha\over c}{\bf E}\times{\bf H}\Bigr)^2 -{1\over 2}\alpha E^2
-{\mu^{(1)}\over n_p}[n_p-|\Psi({\bf R},t)|^2]\Bigr\} \Psi({\bf R},t),
\end{equation}
where  $\alpha$ has the meaning of $e-h$ pair polarizability,  $\alpha(H)=M_H c^2/H^2$. The equation similar to
(\ref{13}) was proposed in Ref. 17 
on the basis of phenomenological considerations.

\section{VORTICES AND COLLECTIVE EXCITATIONS}

 From (\ref{13}) it is not difficult to
derive the continuity equation $\partial |\Psi|^2/\partial t + {\rm div} {\bf j}_s
=0$, where the superfluid current density can be written as ${\bf j}_s = |\Psi|^2 {\bf
v}_s$. In this case the superfluid velocity ${\bf v}_s$ appears dependent on the
electric field

\begin{equation}\label{p12}
{\bf v}_s = {1\over M_H} \Bigl(\hbar\frac{\partial\varphi}{\partial{\bf R}}+ {\alpha\over c}{\bf E}\times{\bf
H}\Bigr)
\end{equation}
Here $\varphi$ is the phase of the wave function of the condensate
($\Psi=|\Psi|e^{i\varphi}$).

The right-hand side of Eq.(\ref{13}) can be considered as a variational derivative ${\delta F/\delta \Psi^*}$,
where $F$ is the functional of the Ginzburg-Landau type. The variation of $F$ in the electric field gives the
dipole moment ${\bf P}$ of the unit area of the bilayer system. Using of expression (\ref{p12}) for the
superfluid velocity one can obtain the following simple expression for ${\bf P}$:

\begin{equation}\label{14}
{\bf P} = \alpha \Bigl({\bf E}+ {1\over c}{\bf v}_s\times{\bf H}\Bigr)|\Psi|^2.
\end{equation}
This result means that not only the electric field ${\bf E}$, but
also the Lorentz force polarizes the pair, acting in opposite
directions on the positive and negative charges of the pair.

 Eq.(\ref{13}) has inhomogeneous solutions that
describe the quantized vortices. The steady vortex state of the
bilayer system with vortex centered at points ${\bf R}_n$ is
determined by the relation

\begin{equation}\label{15}
{\rm curl}_z\Bigl(\frac{\partial\varphi}{\partial{\bf R}}\Bigr)=
2\pi\sum_n\sigma_n\delta({\bf R}-{\bf R}_n),
\end{equation}
where $\sigma_n=\pm 1$  is sign of the $n$-th vortex circulation,
and $\delta({\bf R})$ is the two-dimensional $\delta$-function.
The two-dimensional density of the polarization charge
($\rho_{pol}=-{\rm div}{\bf P}$), associated with the vortices,
has the form \cite{15}

\begin{equation}\label{16}
\rho_{pol}=-\alpha n_p \frac{2\pi\hbar H}{M_H c}
\sum_n\sigma_n\delta({\bf R}-{\bf R}_n).
\end{equation}
Hence it follows that vortices have the electric charge  $q_n=\pm e \nu$, where
$\nu=2\pi l_H^2 n_p$.

In the presence of a uniform electric field the dispersion law of
collective excitations of the condensate takes the form

\begin{equation}\label{17}
\omega({\bf k}) = {\bf k \cdot v}_s +\sqrt{\Bigl({{\hbar^2k^2}\over {2M_H}}+ 2\mu^{(1)}\Bigr){k^2\over {2M_H}}},
\end{equation}
where the electric field induced superfluid velocity ${\bf v}_s({\bf E}) = {1\over M_H}{\alpha\over c}{\bf
E}\times{\bf H}$. Eq.(\ref{17}) represents the dispersion law for the excitations propagating in the condensate
that moves at a velocity of ${\bf v}_s({\bf E})$. According to the Landau superfluidity criterion, the $v_s$
value must not exceed the critical value $c_0=\sqrt{\mu^{(1)}/M_H}$. The critical velocity $c_0$ determines the
critical electric field $E_{cr}= (c_0/c)H$. For $n_p \sim {\rm 10^{10} cm^{-2}} $, $d \sim {\rm 10^{-6} cm}$, $H
\sim {\rm 1 T} $, $\varepsilon \sim {\rm 10}$, the estimation gives $E_{cr}\sim {\rm 10^3 V/cm} $.

\section{Bilayer systems with curved conducting layers}

So far we have considered idealized structures, where the conducting layers are assumed to be parallel each
other. In this Section we depart from the idealized situation and take into account a small curvature of the
conducting layers. In the general case, the equations of curved layers may be written as $z=Z_i({\bf r})$, where
the index $i=1(2)$ refers to the electron (hole) layer, and ${\bf r}=(x,y)$ is the two-dimensional
radius-vector. The small curvature of the layer means a small (in comparison with the interlayer distance)
variation of $Z_i({\bf r})$ on the magnetic length $l_H$ scale. Note that the curvature of the layers may arise
accidentally, but it may also be made purposely, as the structure non-ideality leads to a number of rather
interesting properties. In particular, the effective mass of the $e-h$ pair, $M_H$, becomes the function of the
pair coordinates; the regions arise, where the energy of the $e-h$ pair is lowered (exciton traps), and lastly,
$e-h$ pairs can gain the polarization in the $xy$ plane. Below we shall demonstrate that this polarization can
lead to the appearance of quantized vortices even at zero temperature.

The approach developed in previous sections can be easily extended to the case of curved layers. To do this, the
argument $V_{12}$ (i.e., the electron-hole spacing) in Eq.(\ref{7}) should be written in the form $\{[Z_1({\bf
r}_1)-Z_2({\bf r}_2)]^2+({\bf r}_1-{\bf r}_2)^2\}^{1/2}$. As with the derivation of Eq.(\ref{13}), further
transformations come to the transition to the center of mass coordinates of the pair ${\bf R}$ and to the
relative coordinate $\tilde{\bf r}={\bf r}_1-{\bf r}_2$. Making use of the smoothness of functions $Z_i({\bf
r})$, we expand the expressions $Z_{1(2)}({\bf R}\pm \tilde{\bf r}/2)$ in power $\tilde{\bf r}$ to an accuracy
of the second-order terms and then to integrate over the fast variable $\tilde{\bf r}$. As a result, we arrive
at the following equation for the wave function of the condensate:

\begin{eqnarray}\label{41}
i\hbar\frac{\partial\Psi({\bf R},t)}{\partial t} =
\Bigl[\Bigl(i\hbar \frac{\partial}{\partial{\bf R}}-{\alpha({\bf
R})\over c}{\bf E}_{tot}\times{\bf H}\Bigr){1\over 2M_H ({\bf R})}
\Bigl(i\hbar
\frac{\partial}{\partial{\bf R}}-{\alpha({\bf R})\over c}{\bf E}_{tot}\times{\bf H}\Bigr) \nonumber \\
 - {1\over 2}\alpha({\bf R}) E_{tot}^2 + U({\bf R})-\mu +g({\bf R})|\Psi({\bf R},t)|^2\Bigr] \Psi({\bf R},t).
\end{eqnarray}
Here the effective mass $M_H$ of the $e-h$ pair is determined by expression (11), where the interlayer distance
depends on the coordinate of the pair: $d({\bf R})=Z_1({\bf R})-Z_2({\bf R})$. In this case, the polarizability
of the $e-h$ pair $\alpha({\bf R})=M_H ({\bf R})c^2/H^2$ is also dependent on the coordinates.

The vector ${\bf E}_{tot}={\bf E}+{\bf E}_{int}$ where ${\bf E}$ is the external electric field  and ${\bf
E}_{int}$ is an ''internal'' field which, just as ${\bf E}$ does, leads to the $e-h$ pair polarization in the
$xy$ plane. The field ${\bf E}_{int}$ is a consequence of the curvature of the conducting layers and is given by
linear in $\tilde{\bf r}$ terms of the expansion of the functions $Z_1({\bf R}+\tilde{\bf r}/2)$ and $Z_2({\bf
R}-\tilde{\bf r}/2)$:

\begin{equation}\label{42}
{\bf E}_{int}= {e\over {\epsilon l_H^2}}F(d({\bf
R}))\frac{\partial}{\partial{\bf R}}\bar Z({\bf R}).
\end{equation}
Here we have introduced the notations   $\bar Z({\bf R})=(1/2)[Z_{1}({\bf R})+Z_2({\bf R})]$ and

\begin{equation}\label{43}
F(d)=-\frac{1}{2} -
\sqrt{\frac{\pi}{2}}\frac{l_H}{2}\frac{\partial}{\partial
d}\Bigl[\Bigl(1-{d^2\over l_H^2}\Bigr)\exp\Bigl({d^2\over
2l_H^2}\Bigr) {\rm erfc}\Bigl({d\over \sqrt2 l_H}\Bigr)\Bigr].
\end{equation}

To clarify the physical sense of the field ${\bf E}_{int}$, one
can imagine a bilayer system, in some region of which the
conducting layers are the planes parallel to each other, though
being deviated from the basal $xy$ plane. In this region, the
$e-h$ pair has the dipole moment directed along the normal to the
conducting layers. Since the normal is deviated from the $z$-axis,
the dipole moment acquires the component parallel to the $xy$
plane. The polarization of $e-h$ pairs in the $xy$ plane,
associated with the conducting layer curvature, can be described
through introduction of the internal field ${\bf E}_{int}$
according to Eq.(\ref{42}).

The $U({\bf R})$ term in Eq.(\ref{41}) is similar to the potential energy of the boson in the Gross-Pitaevskii
equation and is determined  by  the pair binding energy: $U({\bf R})=E_0(d({\bf R}))$. We neglect the
corrections to $U({\bf R})$ connecting with the finite size of the $e-h$ pair, because they are small as the
squared ratio of $l_H$ to the characteristic size of the curvature along the $xy$ plane.  As concerns the last
two terms in Eq.(\ref{41}), we note that the chemical potential $\mu$ must be derived from the condition of
normalization of $\int |\Psi|^2d^2{\bf R}$ to the total number of $e-h$ pairs, while the coordinate-dependent
coupling constant  is  $g({\bf R})=\mu^{(1)}(d({\bf R}))/n_p$, where the function $\mu^{(1)}(d)$ is determined
by Eq.(\ref{11}).

It can readily be demonstrated that Eq.(\ref{41}) leads to the continuity equation $\partial n/\partial t + {\rm
div}(n {\bf v}_s) =0$, where the condensate density $n= |\Psi({\bf R},t)|^2$, and the superfluid velocity is
given by:

\begin{equation}\label{44}
{\bf v}_s = {1\over M_H({\bf R})} \Bigl(\hbar\frac{\partial\varphi}{\partial{\bf R}}+ {\alpha({\bf R})\over
c}{\bf E}_{tot}\times{\bf H}\Bigr).
\end{equation}

In an inhomogeneous bilayer system, where the effective mass of the pair is dependent on its  coordinates, it is
convenient to pass from the description in terms of superfluid velocity to the description in terms of the
generalized pair momentum ${\bf P}_s  = \hbar\frac{\partial\varphi}{\partial{\bf R}}$. For example, the dynamic
equation for the phase of the order parameter, which follows directly from Eq.(\ref{41}), can be written in
terms of the generalized momentum of the pair as

\begin{equation}\label{45}
\frac{\partial {\bf P}_s}{\partial{t}} = -\frac{\partial}{\partial{\bf R}}\Bigl[\frac{1}{2 M_H}{\bf P}_s^2
+\frac{\alpha}{M_H c}{\bf P}_s\cdot {\bf E}_{tot}\times{\bf H}
-\frac{\hbar^2}{2\sqrt{n}}\frac{\partial}{\partial{\bf
R}}\Bigl(\frac{1}{M_H}\frac{\partial{\sqrt{n}}}{\partial{\bf R}}\Bigr)+U-\mu +g n\Bigl].
\end{equation}

One can see from Eq.(\ref{44}) that the difference of the generalized momentum $\bf{P}_s$ from the kinematic
momentum ${\bf p}_s=M_H {\bf v}_s$ is connected with the  vector ${\bf E}_{tot}\times{\bf H}$. An important
point is that  the circulation of the generalized momentum  is quantized, while in the general case the
circulation of ${\bf p}_s$ is not quantized. In particular, for the singly connected vortex-free region ${\rm
curl}\,{\bf P}_s=0$, and the kinematic momentum satisfies the equation ${\rm curl}\,{\bf p}_s=-{\frac{\bf
H}{c}}\,{\rm div}\,[\alpha({\bf R}) {\bf E}_{tot}]$. In other words, even in the absence of quantized vortices,
the vector ${\bf p}_s$ may have a vortex component because of the inhomogeneous condensate polarization
$\alpha({\bf R}) {\bf E}_{tot}$. Note that in the homogeneous system, where ${\bf E}_{tot} = {\bf E}$, the last
equality reduces to the known condition  ${\rm curl}\,{\bf v}_s=0$  if ${\rm div}\, {\bf E}=0$. Thus the
condition ${\rm curl}\,{\bf v}_s=0$ is a special case of the more general quantum condition $\oint {\bf P}_s
d{\bf R} = 2\pi\hbar k$, where $k$ is an integer.

The dipole moment ${\bf P}$ of the unit area of the inhomogeneous
bilayer system is given by expression (17), where ${\bf E}$ must
be replaced by ${\bf E}_{tot}$. Calculating ${\rm div}\,{\bf P}$
we obtain the following expression for the two-dimensional
polarization charge density:

\begin{equation}\label{46}
\rho_{pol}=-{\frac{ c n}{H}}\,{\rm curl}_z{\bf P}_s-{\frac{ c }{H}}\,\Bigl(\frac{\partial n}{\partial{\bf
R}}\times {\bf P}_s\Bigr)_z.
\end{equation}
Eq.(\ref{46}) can be used to discuss the peculiarities of vortices in inhomogeneous bilayer systems. We restrict
the discussion to two cases. First we consider the vortex  at the center of the axially symmetrical trap for
$e-h$ pairs. From Eq.(\ref{46}) it follows that for this vortex the polarization charge density is given by

\begin{equation}\label{47}
\rho_{pol}=-2\pi \sigma e l_H^2\, n(0) \,\delta({\bf R})-\sigma e
l_H^2{\frac{1}{R}}{\frac{d n}{d R}}.
\end{equation}
Here $\sigma = \pm 1$ is the sign of the vortex circulation. By
integrating Eq.(\ref{47}) over ${\bf R}$ we obtain that the charge
concentrated in the region of radius $R$ is determined by the
$e-h$ pair density at the boundary of the region: $q(R)=-2\pi
\sigma e l_H^2\, n(R)$. This expression transforms to the formula
derived in the previous section for the coordinate-independent
vortex charge. Indeed, in the homogeneous system, far from the
vortex core, the density $n(R)$ quickly approaches the constant
$n_p$ value. Therefore, at $R\gg \xi$ the vortex charge is
independent of $R$ and is equal to $q=-\sigma e \nu$. On the other
hand, if the vortex is located at the center of the trap, where
the captured $e-h$ pairs are contained in the $R<R_0$ region, then
the total electric charge associated with the vortex equals zero
(since $n(R)=0$ at $R>R_0$). The contradiction with the previous
result is removed if one takes into account that in the
homogeneous system, along with the vortex charge $q$, there arises
the opposite charge $(-q)$ at the boundaries of the system.

To conclude this Section, we discuss the question whether the curvature of conducting layers may give rise to
vortices. For simplicity, we consider the axially symmetrical situation, $\bar Z ={\bar Z}(R)$, and put
$d(R)={\rm const}$. In this case, we have the effective mass of the $e-h$ pair $M_H={\rm const}$, the field
${\bf E}_{tot}={\bf E}_{int}= {e\over {\epsilon l_H^2}}F(d)\frac{\partial {\bar Z}}{\partial R}{\bf e}_r$, and
the superfluid velocity ${\bf v}_s = {1\over M_H} \Bigl(\frac{\hbar k}{ R}- \frac{M_H e^2}{\epsilon \hbar}
F(d)\frac{\partial {\bar Z}}{\partial R}\Bigr){\bf e}_\theta$. Here $k$ is the integer, ${\bf e}_r$ and ${\bf
e}_\theta$ are, respectively, the radial and azimuthal unit vectors. The $k$-dependent contribution to the
energy of condensate  is given by the integral $\int (M_H/2)v_s^2|\Psi|^2d^2{\bf R}$. At $d\ll l_H$ the
condensate energy is minimal when

\begin{equation}\label{48}
k={\rm Int}\Bigl(\frac{d[\bar Z (R_0)-\bar Z (0)]}{2 l_H^2
\log(R_0/\xi)}\Bigr),
\end{equation}
where $R_0$ is the radius of the system, and the function ${\rm Int}(x)$ defines the integer closest to $x$. As
follows from Eq.(\ref{48}),  the fulfillment of a rather rigid condition $d|\bar Z (R_0)-\bar Z (0)|>l_H^2
\log(R_0/\xi)$ gives  $k\neq 0$, i.e., the curvature of layers may give rise to quantized vortices even at zero
temperature.

\section{CONCLUDING REMARKS}

In conclusion, we justify, in short, the mean-field approximation used above.  As was shown for the first time
by Shick \cite{16}, in a two-dimensional  Bose system in the low-density limit the interaction is renormalized
essentially with respect to the bare one. In particular, for the two-dimensional bosons modeled as hard dicks of
the radius $a$ the summation of the infinite series of the ladder diagrams yields the effective interaction in
the form $4\pi \hbar^2/m \ln (1/na^2)$. One can think that for the electron-hole pairs considered a similar
logarithmic dependence of the interaction on the density of bosons would take place as well. In reality, at
$d\ll l_H$ it is not the case.

The validity of the mean-field approximation at $d\ll l_H$ can be
justified as follows. In rarefied systems the effective
interaction is proportional, in the leading order, to the
amplitude of scattering $f$ of two bosons on each other. For the
bosons with the wavelength of order of the average interparticle
distance $n^{-1/2}$ in the case of hard-disk scattering the
amplitude of scattering is $f\sim (\ln 1/na^2)^{-1}$ and we arrive
at the Shick's result. It is important to emphasize that in
deriving of that result disks considered as impenetrable ones. But
the electron-hole pairs may overlap and it is better to consider
them as "penetrable" discs. If one models the pairs as discs with
the interaction potential equal $U=U_0$ inside the discs and $U=0$
outside the discs one can easily obtain the scattering amplitude
\begin{equation}\label{18}
   f=\frac{2\pi}{\frac{\hbar^2}{ma^2 U_0}+\ln \frac{1}{na^2}}
\end{equation}
(with the logarithmic accuracy). The first term in the denominator can be evaluated from Eq.(\ref{11}) for the
chemical potential. At $d\ll l_H$ for the radius of the disc $a=l_H$ this term is of order of $l_H^2/(4\pi
d^2)$. Substituting the density in the form $\nu/2\pi l_H^2$ and again using $l_H$ as the radius of the disc $a$
we obtain the second term in denominator in the form $\ln 2\pi/\nu$. In real physical systems $\nu\gtrsim 1/10$
(at smaller $\nu$ the localization effects become important) and the first term is much larger than the second
one. Therefore, the effective interaction is proportional $U_0$ and coincides with the mean-field result. Thus,
at $d\ll l_H$ the mean-field approximation is self-consistent and its application is well founded.

    This work was supported by the INTAS grant No 01-2344.

\end{document}